\documentclass[useAMS,usenatbib,usegraphicx]{mn2e}
\usepackage{subfigure}      

\title[Core instability models of giant planet accretion II]{Core instability models of giant planet accretion II:\\ forming planetary systems}
\author[Y. Miguel and A. Brunini]{Y. Miguel$^{1,2}$\thanks{E-mail: ymiguel@fcaglp.unlp.edu.ar} and A. Brunini$^{1,2}$\thanks{Member of the Carrera del Investigador Cient\'\i fico. Consejo Nacional de Investigaciones Cient\'\i ficas y T\'ecnicas (CONICET).E-mail: abrunini@fcaglp.unlp.edu.ar}\\
$^1$Facultad de Ciencias Astron\'omicas y Geof\'\i sicas. Universidad
Nacional de La Plata. 1900 La Plata, ARGENTINA.\\
$^2$Instituto de Astrof\'\i sica de La Plata (CCT La Plata-CONICET, UNLP), Paseo del Bosque S/N, La Plata, Argentina}

\begin{document}

\pagerange{\pageref{firstpage}--\pageref{lastpage}}

\label{firstpage}

\maketitle
        
\begin{abstract}
      
We develop a simple model for computing planetary formation based on the core instability model for the gas accretion and the oligarchic growth regime for the accretion of the solid core. In this model several planets can form simultaneously in the disc, a fact that has important implications specially for the changes in the dynamic of the planetesimals and the growth of the cores since we consider the collision between them as a source of potential growth. The type I and II migration of the embryos and the migration of the planetesimals due to the interaction with the disc of gas are also taken into account. With this model we consider different initial conditions to generate a variety of planetary systems and analyse them statistically. We explore the effects of using different type I migration rates on the final number of planets formed per planetary system such as on the distribution of masses and semimajor axis of extrasolar planets, where we also analyse the implications of considering different gas accretion rates. A particularly interesting result is the generation of a larger population of habitable planets when the gas accretion rate and type I migration are slower.
\end{abstract}


\section{Introduction}

Planetary astronomy is a young science, and until recently, was essentially devoted to the study of planetary bodies in our own Solar System. The extrasolar planets found since 1995 have vastly expanded our database by increasing the number of known planets by more than 200. Although the distribution of masses and semi-major axis of observed extrasolar planets is highly biased towards those planets that are detectable using Doppler radial velocity technique, the increase in precision and continuity of the surveys has given the possibility of unveiling a large variety of planets and reformulate the theories of planetary formation. As the number of detections grows, statistical studies of the properties of exoplanets and their host stars can be conducted to unravel some of the processes leading to the formation of planetary systems.

In this frame several models of planetary formation have been presented in the last years. \citet{b4} and \citet{b33} have developed semi analytical models based on the oligarchic growth regime, that retain much of the simplicity of the analytic models, but incorporate more physics from the detailed ones. \citet{b11,b12,b13} have studied the mass and semimajor axis distribution of extrasolar planets through a very simple model but considering all the relevant physics involved in the process of planetary formation.

We begin our studies in a previous paper \citep{b26}, where we developed a very simple model based on the core instability model. The accretion of solids was based in the regime where the largest embryos dominate the dynamics of the planetesimal disc: the oligarchic growth regime \citep{b19,b14}. When the cores are large enough to have an associate envelope which can not be sustained by the hydrostatic equilibrium any more, the gas accretion process begins. In our previous work we explored the effects of different gas accretion rates on the final mass distribution of extrasolar planets, and found a strong dependence. But our simple model allowed us to form only one core per disc and did not allow us to consider the dynamical evolution of the cores or the planetesimal disc. 

In this paper, we have improved the model introduced in Paper I considering other important effects. The mainly improvements are the formation and evolution of several cores simultaneously in the protoplanetary nebula, a fact that also influences the dynamic of the planetesimal disc. The embryos can grow by the accretion of planetesimals or by accreting other embryos, in this work we consider the collision among them as a source of growth.

We also incorporate the orbital evolution of the protoplanets due to its tidal interaction with the gas in the surrounding protoplanetary nebula, considering two regimes of planetary migration: type I and II regimes, which result mostly in a radial motion towards the host star. Type I migration involves low mass protoplanets who are rapidly moved towards the star \citep{b9,b35} and type II is the regime of migration for those protoplanets whose mass is large enough to allow them to open up a gap in its orbit and migrate on a rate controlled by the disc viscous time-scale \citep{b22}. This processes could have important consequences on the final mass and semimajor axis distribution, such as on the final number of planets formed per protoplanetary disc.

The disc of planetesimals also interacts with the nebular gas. This gas drag effect causes the migration of planetesimals towards the host star before they become large enough to decouple from the disc gas. This effect, which was included in our model, affects directly the distribution of solids in the disc and the growth of the cores.    
        
We show that a variety of planetary systems can be constructed using these principles, varying the initial conditions, and that the mass and semimajor distribution of extrasolar planets is strongly dependent on the gas accretion model and type I migration rate considered, being larger the population of habitable planets when the rate of gas accretion and type I migration are slower.

\section{Description of the model}

In our previous work \citep{b26}, Paper I, we developed a simple model for computing planetary formation based in a work of \citet{b11}(hereafter IL04), but incorporating the accretion rates given by \citet{b6}. This lead us to a very different results regarding the final mass distribution of extrasolar planets, which was found strongly dependent on the gas accretion model considered.

With the aim of improve the model introduced in Paper I, we have extended it by considering other important effects. In this section we explain briefly our previous model and show the improvements made.

\subsection{The previous model}\label{modelo-anterior}

For the sake of completeness, we will explain our previous model. Although very  simple, it incorporates all the important aspects of planetary formation. 

We considered one initial core per protoplanetary disc, with an initial mass of $10^{-5}M_{\oplus}$. The initial disc was based on the minimum mass solar nebula of \citet{b10}(for more details see section \ref{nebula}), and it was defined between $0.01$ and $100 au$.  The population of the nebular mass scaling parameter, $f_d$, which states the solid mass in the disc in terms of the minimum mass solar nebula model, was distributed with a Gaussian distribution in terms of $log_{10}(f_d)$ with dispersion 1 and centred at 0.25, in order to characterise different discs. We supposed that $f_g$, which is a scaling parameter for the gas disc, was equal to $f_d$, this correspond to discs with solar metallicity.

The solid and gaseous disc were not time-invariant. The gaseous disc changed globally, decaying exponentially with a characteristic time-scale of $4x10^6 years$, in accordance to current estimates of disc lifetimes around young solar type stars \citep{b2} and the solid disc changed locally, suffering the depletion of planetesimals produced by the effect of core's accretion. 

The $10^{-5} M_{\oplus}$ initial core was located with equal probability per interval of $log(a)$, and it grew with the following solids accretion rate based on the particle-in-a-box-approximation \citep{b29}, 

\begin{equation}\label{coreaccretion}
\frac{dM_c}{dt}=10.53 \Sigma_d \Omega_K R^2 \Big(1+\frac{2GM_t}{R \sigma}\Big)
\end{equation}  
where R is the physical radius of the core, $\Omega_K$ is the Kepler frequency, $M_t$ is the total mass of the core, $\Sigma_d$ is the solids disc surface density and $\sigma$ is the relative velocity between the embryo and the disc of planetesimals. 

The factor $10.53$ was introduced for considering the $F$ factor introduced by \citet{b8}, and the approximations made for the eccentricity $e$, the inclination $i$ and the disc scale of high $h(a)$ in the high-$\sigma$ equilibrium regime, $2i \simeq e$ and $h(a) \simeq ai$.

Finally, a factor of 4 was introduced in the above equation, which is an approximate value taken in order to fit the solid accretion rate used by \citet{b6}, which includes the  evolution of the planetesimals r.m.s. $e$ and $i$ and the drag effect caused by the gaseous envelope on the incoming planetesimals. When the gravitational perturbation due to the protoplanets is balanced by dissipation due to the gas drag, the planetesimals r.m.s. eccentricity attains an equilibrium value, which was obtained by \citet{b33}. This effect was considered in the simulations performed by \citet{b6}, where the solid accretion rate was prescribed as that obtained by \citet{b33}, and where the enhancement of the protoplanet's cross section due to the drag effect caused by the protoplanet's gaseous envelope on the entering planetesimals was also considered. We compared the core accretion rates found without considering these effects with all the cases considered by \citet{b6}. The results showed that the core accretion rate is $\sim 4$ times larger when the evolution of the $e$ and $i$ are considered \citep{b26}. 

When a core of mass $M_c$ reached a certain critical mass, it started the gas accretion on to the core. The critical mass at which the gas accretion process began was obtained analytically by \citet{b30} and it was improved by \citet{b16} through numerical simulations. We considered a simplified version, 

\begin{equation} \label{mcritica}
M_{crit}\sim \bigg(\frac{\dot{M_c}}{10^{-6}M_{\oplus}yr^{-1}}\bigg)^{\frac{1}{4}}
\end{equation} 

The gas accretion rate was obtained by fitting the results of the self-consistent code developed by \citet{b6} 

\begin{equation} \label{acregas1}
\frac{dM_g}{dt}=\frac{M_t}{\tau_g}
\end{equation}
where $M_g$ is the mass of the surrounding envelope and $\tau_g$ is its characteristic growth time, 

\begin{equation} \label{acregas2}
\tau_g=1.64x10^9 \bigg(\frac{M_t}{M_{\oplus}}\bigg)^{-1.91}yrs
\end{equation}

This process stopped when the protoplanet consumed all the gas available on its feeding zone, or what is the same, when it reached its isolated gas mass, given by,

\begin{equation}
M_{g,iso}=2\pi a b r_H \Sigma_g
\end{equation}
where $\Sigma_g$ is the gas surface density and $br_H$ is the typical distance between two adjacent embryos, which according to \citet{b19} is $\sim 10r_H$,  being $r_H$ the Hill radius.

\subsection{The protoplanetary nebula and the initial cores}\label{nebula}

As in our previous paper \citep{b26}, the structure of the protoplanetary nebula is based on the minimum mass solar nebula (MMSN) of \citet{b10}, where the surface density of solids is defined as follows:

\begin{equation}
\Sigma_d=7 f_d \eta_{ice} \bigg(\frac{a}{1au}\bigg)^{-\frac{3}{2}} gcm^{-2}
\end{equation}
where $\eta_{ice}$ is $1$ inside the ice condensation radius and $4$ outside it, expressing the effect of water ice formation across the ice boundary. The ``snow line'' is located at $a_{ice}=2.7\big(\frac{M_{\star}}{M_{\odot}}\big)^2$ $au$ from the central star of mass $M_{\star}$.

Similarly, the volume density of gas follows an exponential profile, as explained \citet{b33}, which is

\begin{equation}
\rho(a,z)=\rho_{g,0}(a)e^{\frac{z^2}{h(a)^2}}  gcm^{-3}
\end{equation}
where $\rho_{g,0}$ is the mid-plane value density given by,
 
\begin{equation}
\rho_{g,0}(a)=1.4x10^{-9} f_g \bigg(\frac{a}{1au}\bigg)^{-\frac{11}{4}}  gcm^{-3}
\end{equation}

Unlike the previous calculations, we assume that the surface density of gas {\it everywhere} varies with time as:

\begin{equation}\label{gasdecay}
\rho_g=\rho_{g,0} e^{-\frac{t}{\tau_{disc}}}
\end{equation}
with $\tau_{disc}$ the gas depletion time-scale.

We consider a disc between $R_{in}$ and $30$ $au$, where the inner disc radius is estimated by,

\begin{equation}\label{innerradius}
R_{in}=0.0344 \Psi \bigg(\frac{1500^{\circ}K}{T_{sub}}\bigg)^2\bigg(\frac{L_{\star}}{L_{\odot}}\bigg)^{\frac{1}{2}} au
\end{equation}
as explained \citet{b34}. In the above equation $\Psi$ is a factor of $\approx 2$ which depends on the disc structure and radiative transfer, the dust sublimation temperature, $T_{sub}$, is taken as $1500^{\circ}K$ and $L_{\star}$ and $L_{\odot}$ are the stellar and Sun luminosity respectively.

In this work we consider stars with masses between $0.7$ and $1.4 M_{\odot}$, the stellar luminosity to this range of masses is given by

\begin{equation}
\frac{L_{\star}}{L_{\odot}} \simeq \bigg(\frac{M_{\star}}{M_{\odot}}\bigg)^{4}
\end{equation}

In our previous work we considered one core per disc, here we form several cores in the same disc. Initially we start the simulation with a number $N_{planets,0}$ of cores through the disc, separated by $10 r_H$. This could have important consequences on the final distribution of masses and semimajor axis of extrasolar planets, specially for the changes in the dynamic of the planetesimals, and also considering the collisions between embryos as a very important source of growth, as was shown by \citet{b3}.

In the core instability model the first step in the planetary formation is the coagulation of dust particles. When particles of many meters in size have been formed, the gravitational forces became more important than gas drag, and the larger planetesimals grow faster than the smaller ones. This stage of increasingly rapid growth is the runaway growth \citep{b7,b18}. In this stage, the planetesimal dynamic is dominated by the interaction with other planetesimals, until the larger ones grows so much and start to dominate the velocity distribution of the planetesimals, \citep{b19}. This is the oligarchic growth stage, and is the stage where the embryos form. In this high-$\sigma$ regime, collisions between planetesimals tend to be destructive leaving fragments of planetesimals as a result of the impacts. This fragments are small enough to be rapidly damped by the gas drag and as a consequence, gravitational focusing favours the embryo, increasing its accretion rate \citep{b4}. This could be an important factor in the growth of the embryos, but is not taken into account in this work.

\citet{b14} found through N-body simulations, the minimum mass necessary for starting the oligarchic growth stage, which is given by   

\begin{equation}\label{Masa-inicial}
M_{oli}\simeq \frac{1.6 a^{\frac{6}{5}}b^{\frac{3}{5}}m^{\frac{3}{5}}\Sigma_d^{\frac{3}{5}}}{M_{\star}^{\frac{1}{5}}}
\end{equation}
where m is the effective planetesimal mass, which  was found adopting the next a cumulative power law mass distribution of planetesimals,

\begin{equation}
N(m)dm \propto m^{-p}dm
\end{equation} 
where $p$ is $1.8-3$. We consider $p=2.5$,consistent with the results of \citet{b19,b20} for the spectrum of masses of a population that has relaxed to isolated runaway bodies. Then the typical mass we use is found through

\begin{equation}\label{planetetipica}
  m=\frac{\int_{m_{min}}^{m_{max}}m' N(m')dm'}{\int_{m_{min}}^{m_{max}} N(m')dm'}
\end{equation}
here we suppose $m_{min}=6.3x10^{12}g$ and $m_{max}=6.3x10^{21}g$ are the minimum and maximum planetesimal mass, equivalent to $0.1-100 Km$ radius (e.g. the same values adopted by \citet{b3}), here we note that the disc mass is contained in the small bodies. The typical planetesimal mass found with equation (\ref{planetetipica}) is the one used and is a constant in all the simulations performed.

One initial core is located at $a=R_{in}$, the rest of the cores are separated $10$ $r_H$ each other until the end of the disc is reached. Their initial masses are given by equation (\ref{Masa-inicial}), where we see that different location in the disc leads to a different initial oligarchic mass.

\subsection{The growth of the cores}\label{tasas-de-acrecion}

The solid accretion rate for a core in the oligarchic growth regime, considering the particle-in-a-box approximation \citep{b29} is given by equation (\ref{coreaccretion}). \citet{b33} found a expression considering the evolution of the planetesimal rms $e$ and $i$. They found that when the embryos' gravitational perturbation are balanced by dissipation due to the gas drag, the planetesimal rms $e$ attains an equilibrium value given by,  

\begin{equation}\label{excentricidad}
e_{eq}(\simeq 2i_{eq})\simeq \frac{1.7\cdot m^{\frac{1}{15}} M_t^{\frac{1}{3}} \rho_m^{\frac{2}{15}}}{b^{\frac{1}{5}}C_D^{\frac{1}{5}}\rho_{gas}^{\frac{1}{5}}M_{\star}^{\frac{1}{3}}a^{\frac{1}{5}}}
\end{equation}
where $C_D$ is a dimensionless drag coefficient which is of order unity, and $\rho_m$ is the planetesimal bulk density, which is $\simeq 1.5 gcm^{-3}$. The equation (\ref{coreaccretion}) could be rewritten using this expression, 

\begin{equation}\label{core-accretion}
\frac{dM_c}{dt}\simeq \frac{3.9 b^{\frac{2}{5}}C_D^{\frac{2}{5}}G^{\frac{1}{2}}M_{\star}^{\frac{1}{6}}\rho_{gas}^{\frac{2}{5}}\Sigma_d}{\rho_m^{\frac{4}{15}}\rho_M^{\frac{1}{3}}a^{\frac{1}{10}}m^{\frac{2}{15}}}M_t^\frac{2}{3}
\end{equation}
where $\rho_M$ is the embryo bulk density ($\rho_M\simeq 1.5 gcm^{-3}$).

The main difference between equation (\ref{core-accretion}) and the one considered in Paper I is that in the above equation we take into account the evolution of rms $e$ and $i$ through the equation (\ref{excentricidad}), and in Paper I we approximated this evolution considering the cases analysed by \citet{b6}. 

Equilibrium rms velocities are a good approximation as long as the embryos reach roughly one Earth mass, a state that is quickly attained in the regions $a < 10 au$ \citep{b4}. Nevertheless, our model intent to remain as simple as possible, in such a way that we can perform statistics of a large number of simulated result. Within this in mind, a more appropriate velocity evolution model is for the moment out of our possibilities. 

The growth of the cores terminate when they consume all the planetesimals on their feeding zone ($\Delta a_c= b r_H$), or equivalently when the solid superficial density is $\Sigma_d=0$ (see section \ref{evo-solidos}). But they also can stop the growth when the density of planetesimals is substantially depleted by ejection \citep{b33,b11}, this is, when the ratio of collision to ejection probabilities is 
\begin{equation}
\Big(\frac{v_e}{v_s}\Big)^4 \ll 1
\end{equation} 
where $v_e=\sqrt{2GM_{\star}/a}$ is the escape velocity from the primary and $v_s=\sqrt{GM_t/R}$ is their characteristic surface velocity.

Once the core became massive enough to retain a gas envelope, the effect of this atmospheric gas drag on the planetesimals increases the collision cross section of the protoplanet. Considering the model for a purely radiative atmosphere \citep{b30,b17}, \citet{b4} found an approximate expression for the enhanced collision radius ($R_c$) of the embryo,

\begin{equation}
\bigg(\frac{R_c}{R}\bigg)^4=\frac{0.000344\mu^4cP}{\kappa r_m \Sigma_d}\bigg(\frac{M_t}{M_{\oplus}}\bigg)^2\bigg(\frac{24e_{eq}^2}{24+5e_{eq}^2}\bigg)
\end{equation}
where $c$ is the velocity of light, $P$ is the orbital period of the protoplanet, $\kappa$ is the opacity of the atmosphere which is considered as $\simeq 4 cm^2 g^{-1}$, $r_m$ is the planetesimal's typical radius and the equilibrium eccentricity is considered as $\approx 2$ in this expression. 

Once the core reaches the critical mass given by equation (\ref{mcritica}), the gas accretion process begins. The model for the gas accretion on to the core is the same used in Paper I, and it was already explain in section \ref{modelo-anterior}.

We also perform some simulations considering the gas accretion rate given by IL04, 

\begin{equation}
\tau_{g,(IL04)}=1.x10^9 \bigg(\frac{M_t}{M_{\oplus}}\bigg)^{-3}yrs
\end{equation}

which differs mainly in the exponent, taken as 3 by them. It leads as to a smaller gas accretion rate, which has an enormous influence on the distribution of masses, as was explained on Paper I.

Until this point, we have assumed that embryos grow by accreting gas and planetesimals only.  However, embryos can also grow by accreting other embryos. When two protoplanets are too close to each other, their mutual gravitational perturbation induce high eccentricities, which enable their orbits to cross. This process may lead to close gravitational encounters and violent collisions between the embryos.

We suppose that collision between protoplanets will occur if their orbital spacing is less than 3.5 Hill radius. Considering that the evolution of planetary atmosphere after a merger is a complex and poorly understood process, we analyse two different and extreme scenarios. On the one hand we assume that all embryo-embryo collisions will lead to coalescence to form a single body where the result is simply the sum of the masses (gas and solids) of both embryos, no matter the composition of the protoplanets. On the other hand impacts so energetic can cause a net loss of atmosphere, because the kinetic energy of the impact  is partly transferred to the associate envelope accelerating its molecules to velocities greater than the escape velocity from the protoplanet. So we want to explore this possibility and in the other scenario, if both colliding planets have an associate envelope, after the impact their atmospheres would be lost and the result would be a merger of their cores. The mass and semimajor axis distribution were similar in both cases, for this reason we stayed with scenario 1) to not add greater complexity to the model.

\subsection{Evolution of the solids surface density}\label{evo-solidos}

Protoplanetary discs are not quite steady, so the surface density of solids is not a time-invariant.  It will change over time due to different effects, one of them is the depletion of planetesimals produced by the effect of cores' accretion (this effect was also considered in Paper I). If, $\Sigma_{d,0}$ is the initial surface density, we have to remove what already ate the embryo, so the evolution of $\Sigma_d$ is given by,

\begin{equation}\label{cambio1}
\Sigma_d= \Sigma_{d,0}-\frac{M_c}{2\pi a b r_H}    
\end{equation}

On the other hand, the disc of planetesimals also interacts with the nebular gas. This gas drag effect will cause a radial motion of planetesimals before they become large enough to decouple from the disc gas. The orbital decay occur on a rate \citep{b1,b33},

\begin{equation} 
\frac{da}{dt}\simeq -2 \frac{a}{T_{gas}}\bigg(\frac{5}{8}e^2+\frac{1}{2}i^2+\eta \bigg)^{\frac{1}{2}}\bigg(\eta+ \frac{49}{16}e^2+\frac{1}{8}i^2\bigg) 
\end{equation}
where $T_{gas}$ is the characteristic time-scale of the gas drag and $\eta$ is the fractional deviation of the gas velocity $v_{gas}$ from Keplerian velocity $v_K$ due to the pressure gradient,	

\begin{equation}
\eta= \frac{v_K-v_{gas}}{v_K}=\frac{13 \pi}{64}\Big(\frac{c_s}{v_K}\Big)^2
\end{equation}
with $c_s$ being the sound speed and ${c_s}/{v_K}\simeq h(a)/a$.

The characteristic time-scale of the gas drag is given by,

\begin{equation}
T_{gas}=\frac{m}{(C_D/2)\pi r_m^2 \rho_g a \Omega_K}
\end{equation}

Finally, applying continuity to the planetesimal disc, the gas drag effect acting on the planetesimals leads to a rate of change of the surface density given by,

\begin{equation}\label{cambio2}
\frac{\partial{\Sigma_d}}{\partial{t}}= - \frac{1}{a}\frac{\partial{}}{\partial{a}}\Big(a\Sigma_d \frac{da}{dt}\Big)  
\end{equation}

Equation (\ref{cambio1}) is added to the equation (\ref{cambio2}) as a sink term, so we obtain the total change of the solid surface density in the disc.

\subsection{Model for planetary migration}

An important contribution to our understanding about planetary formation and evolution was the discovery of the firsts Jupiter-like planets orbiting very close to its central star (hot-Jupiters). The difficulties associated with the formation of this objects are most pronounced as we are closer to their host star, and they have awaken an interest in theories for the migration of protoplanetary embryos due to gravitational interaction with the disc. In this section we will consider two regimes of planetary migration: the type I and II regimes, which result mostly in a radial motion towards the star and concern planets of low and high mass respectively.

\subsubsection{type I migration}

  Type I migration acts on low-mass protoplanets, that is, for those embryos who are unable to open a gap on its orbit. Our model for this regime of planetary migration is similar to that of \citet{b13}, who describe in detail the assumptions made. 

This process is caused by an imbalance in the tidal torques from inner and outer disc leading to angular momentum exchange and making the planet drifts relative to the disc material. The time-scales involved in this process are smaller than the discs lifetimes, and were calculated by \citet{b31} through 3D linear simulations,

\begin{displaymath}
\tau_{migI}= -\frac{a}{\dot{a}}=
\end{displaymath}
\begin{displaymath}
=\frac{1}{2.7+1.1 q} \Big(\frac{c_s}{a \Omega_K}\Big)^2 \frac{M_{\star}}{M_p} \Big(\frac{M_{\star}}{2 h \rho_g a^2}\Big)\Omega_K^{-1}
\end{displaymath}
\begin{equation}\label{migI}
\simeq \frac{5.5x10^5}{2.7+1.1q}\frac{1}{f_g}\Big(\frac{M_p}{M_{\oplus}}\Big)^{-1}\big(\frac{a}{1au}\big)^{\frac{3}{2}}\Big(\frac{M_{\star}}{M_{\oplus}}\Big)^{\frac{3}{2}}yrs
\end{equation}
with $q=1-1.5$, we use $1.5$.

The time-scale of this inward migration is smaller than the disc lifetime, since it seems very unlikely, recent works dealing with planet-disc interactions have  proposed mechanisms that could slow down or stop type I migration. Several mechanisms have been proposed like the study of the magneto-hydrodynamic turbulence in the disc \citep{b27}, the effects of a disc containing a toroidal magnetic field \citep{b32}, the tri-dimensional effects studied by \citet{b24}, variation in the temperature gradient and surface density in the disc \citep{b25}, the effect of including a proper energy balance on the interaction of a low-mass planet with a protoplanetary disc \citep{b28}, and so on.

Halting or slowing down the inward migration requires careful computation of tidal effects between the core and the star, this is much beyond the capabilities of our model. For this reason a factor of $\frac{1}{C_{migI}}$ is introduced in equation (\ref{migI}), to consider non linear effects without introduce a mayor degree of complexity to our model. If we want to slow down migration rates, $C_{migI}$ must be $\le 1$. We perform simulations with $C_{migI}=1, 0.1, 0.01$. We also consider that the migration mechanism stops when the core reach the inner edge of the disc given by equation (\ref{innerradius}).

\subsubsection{type II migration}

According to the observed range of orbital semimajor axes, many giant planets are found to a very close proximity to their parent stars. At this distance the equilibrium temperature is above the $170^{\circ} K$, the temperature at which ice (and the icy cores of giant planets) exists. Actually, giant planets would form, preferably, at or outside of the ice line, this implies that they could have suffered great orbital changes after formation.

The model of giant planet type II migration presented here is a very simple one, and is essentially the same use by \citet{b11,b13}, so we will describe it briefly.

 The process starts when the giant planet form a gap in the disc. According to \citet{b23}, a necessary condition to open a gap is

 \begin{equation}\label{gap-condition}
 r_H \geq h
\end{equation}

The above equation leads to a mass condition, when the mass of the protoplanet is larger than

\begin{equation}
M_{gap}=3.75\cdot 10^{-4}\cdot \Big(\frac{a}{1au}\Big)^{\frac{3}{4}} M_{\star}
\end{equation}
a gap is opened in the disc. When this happens, there are tidal forces acting between the protoplanet and the inner and outer edges of the gap, whose balance (or imbalance) depends on their difference in density. The imbalance can result on a migration of the protoplanet 's orbit \citep{b21} inward or outward. The migration time-scale is

\begin{equation}
\tau_{migII}=0.8x 10^6 f_g^{-1}\Big(\frac{M_p}{M_J}\Big)\Big(\frac{M_{\odot}}{M_{\star}}\Big)\Big(\frac{\alpha}{10^{-4}}\Big)^{-1}\Big(\frac{a}{1au}\Big)^{\frac{1}{2}}yrs
\end{equation}  
 
where $\alpha$ is a dimensionless parameter which characterises the viscosity and is taken as $10^{-3}$ through this work.

The protoplanet will migrate towards the star if $a<R_m$ and away if $a>R_m$, where $R_m$ is the radius of maximum viscous couple. This radius depends on the distribution of gas surface density, so it changes because $\Sigma_g$ is not a constant. From the conservation of angular momentum, and considering the total disc mass as a constant, it can be found that    

\begin{equation}
R_m=10 e^{\frac{2t}{5\tau_{disc}}} \hspace{1cm} au
\end{equation}

Various stopping mechanisms have been proposed, but none of them seem to be effective in halting migration, for this reason we stop migration arbitrarily when the planet reach the inner radius of the disc, which is given by equation (\ref{innerradius}).

When a giant planet is migrating inwards, it will perturbate the cores located on its path towards the host star. This encounter can cause the ejection or the accretion of the core by the giant planet, or neither, which means that the core will survive the giant planet path. 
  
The effect of the migrating giant planets on the formation of terrestrial planets was studied by \citet{b5}, who present results of their N-body simulations of giant planet migration through an inner protoplanet/planetesimal disc, where the migration of the low-mass embryos (type I) was also considered.

At the end of the simulations, they found that $\sim 71\%$ of the initial solid disc survive after the migration of the giant planet, a very small percentage is ejected into a hyperbolic orbit (less that $0.2\%$), and the rest is accreted.

We introduce this results in our simulations, with a very simplified model where in a close encounter between a giant migrating inwards and another core, the last one has the $29\%$ of possibilities of being accreted by the giant and the rest $71\%$ of surviving the giant passage with its orbit unchanged. We note that this is a very simple model, but in reality this fraction will change depending on the distance of embryos from the star.

 This effect does not significantly change the results, since we have more than one migrating giant planet in the disc. The passage of many giant planets will reduce the surviving probabilities of the cores, until they were finally accreted.

\section{Results}

In this section we first analyse the characteristics of the planetary systems formed, and then discuss the effects of different prescriptions for the gas accretion rate and different retardation constants for type I migration on the mass and semimajor axis distribution of extrasolar planets.

\subsection{Characteristics of the planetary systems formed}

In each simulation we generate 1000 discs. Each system evolves for $10^7$ years, in a rich-metal disc, where the time-scale for the depletion of the gas, has a uniform log distribution between $10^6$ and $10^7$ years, and the stellar mass has a uniform distribution in log scale in the range of $0.7-1.4M_{\odot}$. The initial number of planets per disc, $N_{planets,0}$, depends on the initial cores' mass (equation \ref{Masa-inicial}), the inner disc radius (equation \ref{innerradius}) and the mass of the host star. The final number of planets, $N_{planets}$, is a result and shows the evolution of the planetary system.

\begin{figure}
  \begin{center}
    \subfigure[]{\label{fig:nf-c1}\includegraphics[angle=270,width=.4\textwidth]{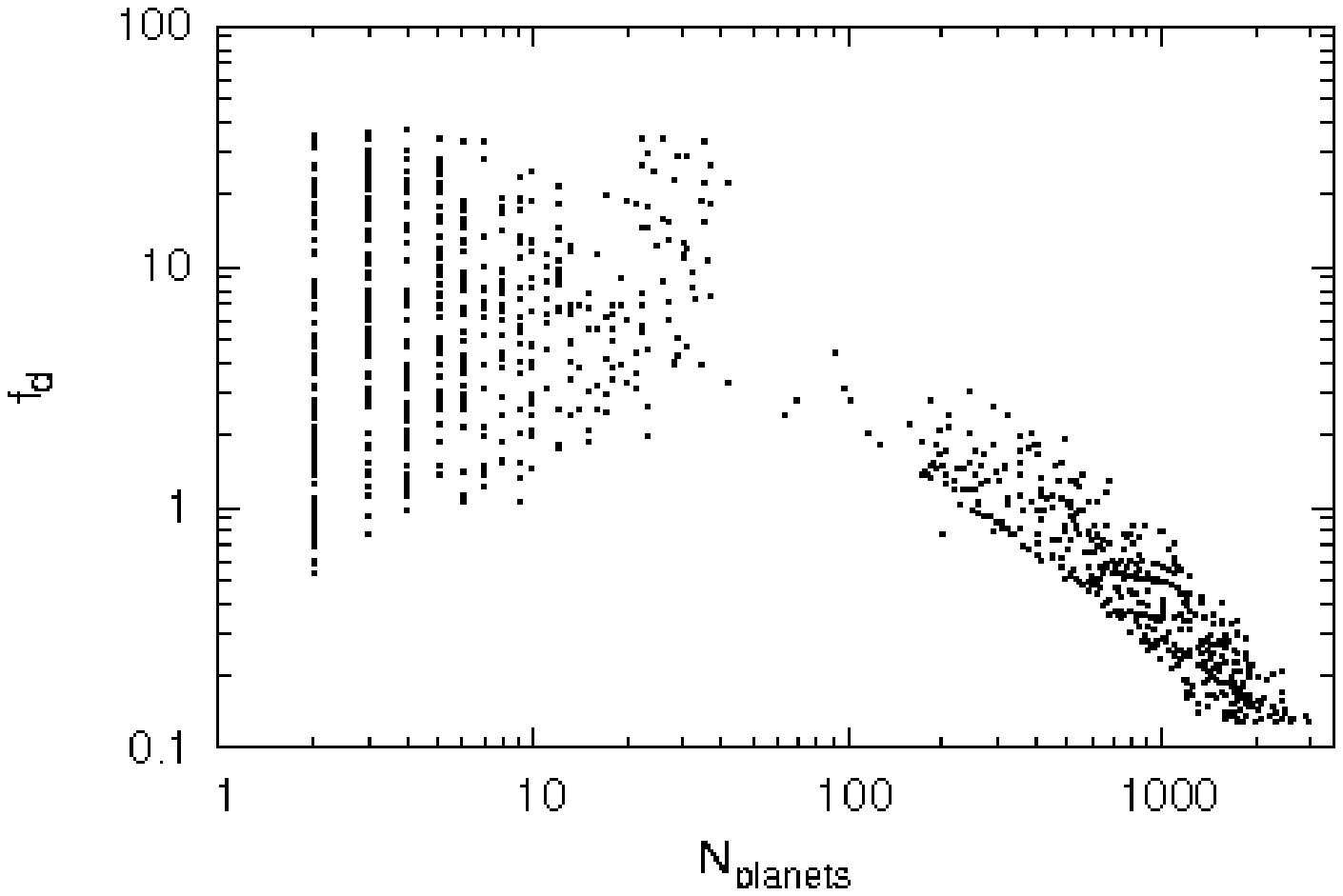}}
    \subfigure[]{\label{fig:nf-c01}\includegraphics[angle=270,width=.4\textwidth]{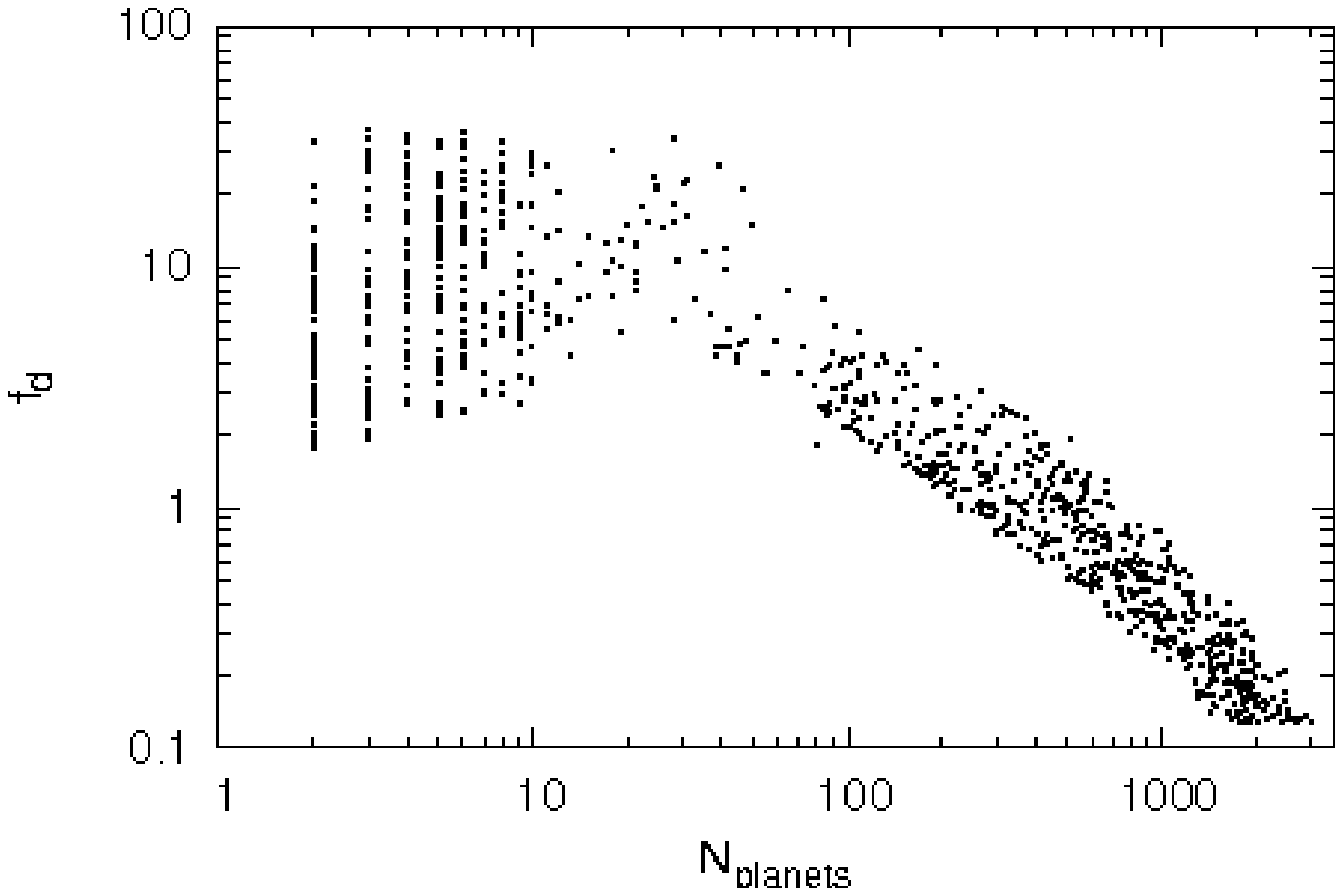}} 
    \subfigure[]{\label{fig:nf-c001}\includegraphics[angle=270,width=.4\textwidth]{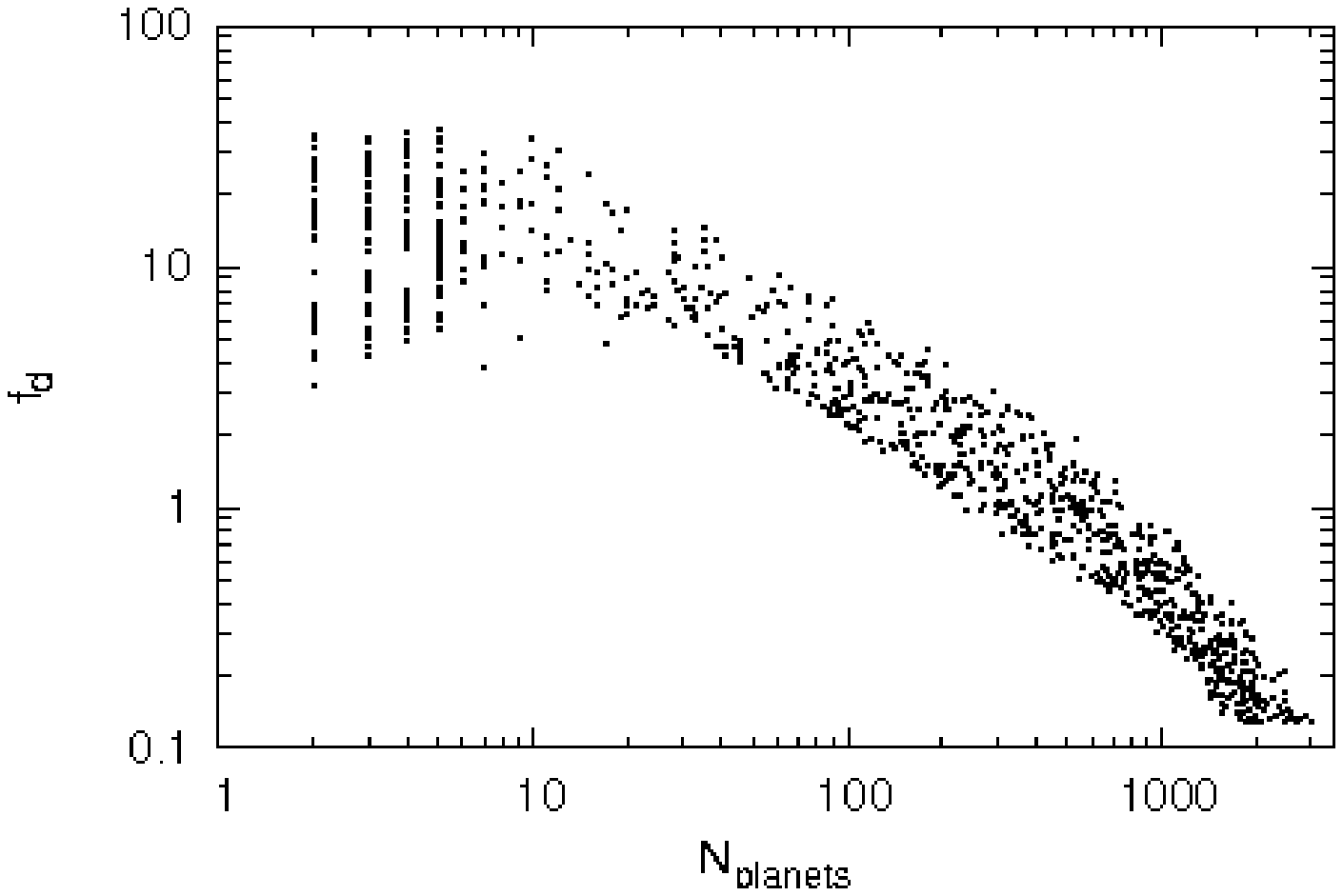}}
  \subfigure[]{\label{fig:nf-c0}\includegraphics[angle=270,width=.4\textwidth]{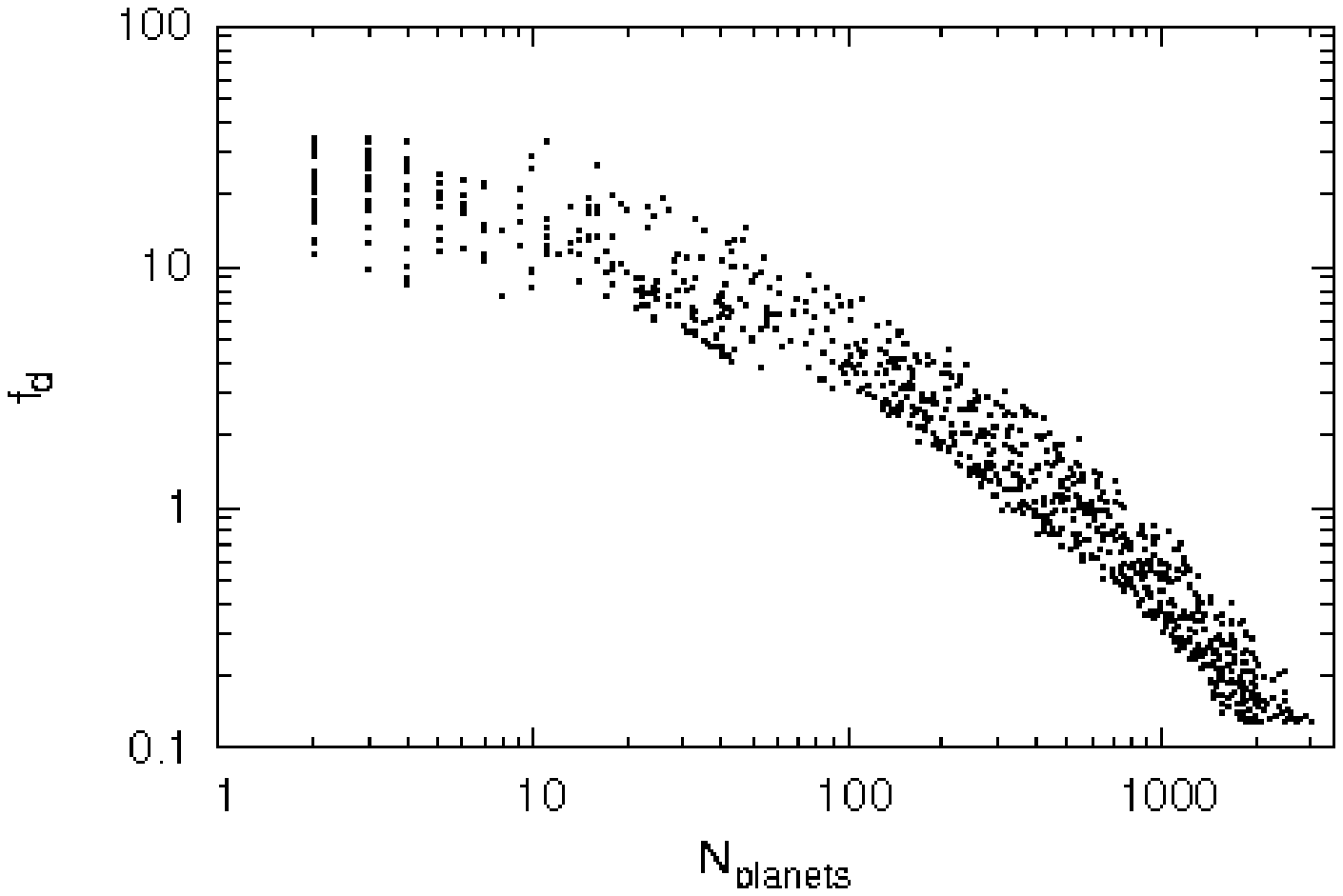}}
  \end{center}
\caption{The final number of planets is plot against the disc mass scaling parameter, $f_d$. Fig. \ref{fig:nf-c1} shows the results when $C_{migI}=1$ is considered, fig. \ref{fig:nf-c01} and \ref{fig:nf-c001} present the results when $C_{migI}$ is equal to $0.1$ and $0.01$ respectively and in Fig. \ref{fig:nf-c0} $C_{migI}$ is $0$ which means that type I migration is not considered. As seen in the figures, planetary systems with a large number of planets are those with small values of $f_d$ and a large disc mass is necessary to form planetary systems with few planets, but the range depends on the type I migration rate considered.}
\label{nf}
\end{figure}

Each disc of gas is defined by $f_{g,0}$, which has a log normal distribution with a dispersion of 1 and centred on $f_{g,0}=1$, and the disc of solids is taken as $f_d=10^{0.1}f_{g,0}$ \citep{b12}. This is slightly different from what we had considered before, because in Paper I we used discs with solar metallicity and here we suppose more metallic discs. The final number of planets per disc also depends on $f_{g,0}$ and $f_d$, as seen in Fig. \ref{nf}, where $f_d$ is plot against $N_{planets}$, considering different values for $C_{migI}$.
  
Figure \ref{fig:nf-c1} was obtained with $C_{migI}=1$, Fig. \ref{fig:nf-c01} considers $C_{migI}=0.1$, the value taken in Fig. \ref{fig:nf-c001} was $0.01$ and the type I migration was not considered in Fig. \ref{fig:nf-c0}.

As seen in the Figures, planetary systems with a large number of planets ($N_{planets}>100$) correspond to small values of $f_d$ ($<3$), and this fact does not change considering different rates of type I migration. These discs have low mass and as a consequence the initial cores will not grow too much. When the mass of the cores is less than $\approx 0.1M_{\oplus}$, they are not so affected by migration (which would favor the merging) and therefore the discs remain with many small planets, no matter the rate of migration considered.

 On the other hand the values of $f_d$ required to form planetary systems with few planets change when different migration rates are considered. When $C_{migI}=1$, discs with $f_d>0.5$ can form planetary systems with few planets. This can be explained with the rapid type I migration rate, which moves the biggest cores inwards accreting the other cores to its path, forming planetary systems with few planets near to the host star. Small values of $f_d$ implies discs less massive, which will form planetary systems with few (terrestrial) planets. 
 
As migration slows down, the range of values of $f_d$ required to obtain planetary systems with few planets becomes smaller, as seen in figures \ref{fig:nf-c01}, \ref{fig:nf-c001} and \ref{fig:nf-c0}. In the last figure, when $C_{migI}=0$, we note that those planetary systems with few planets correspond to $f_d>10$, which leads to high mass discs which will generate giant planets who accrete all the other cores to its path towards the central star.

\begin{figure}
  \begin{center}
    \subfigure[]{\label{fig:histon-c1}\includegraphics[angle=270,width=.44\textwidth]{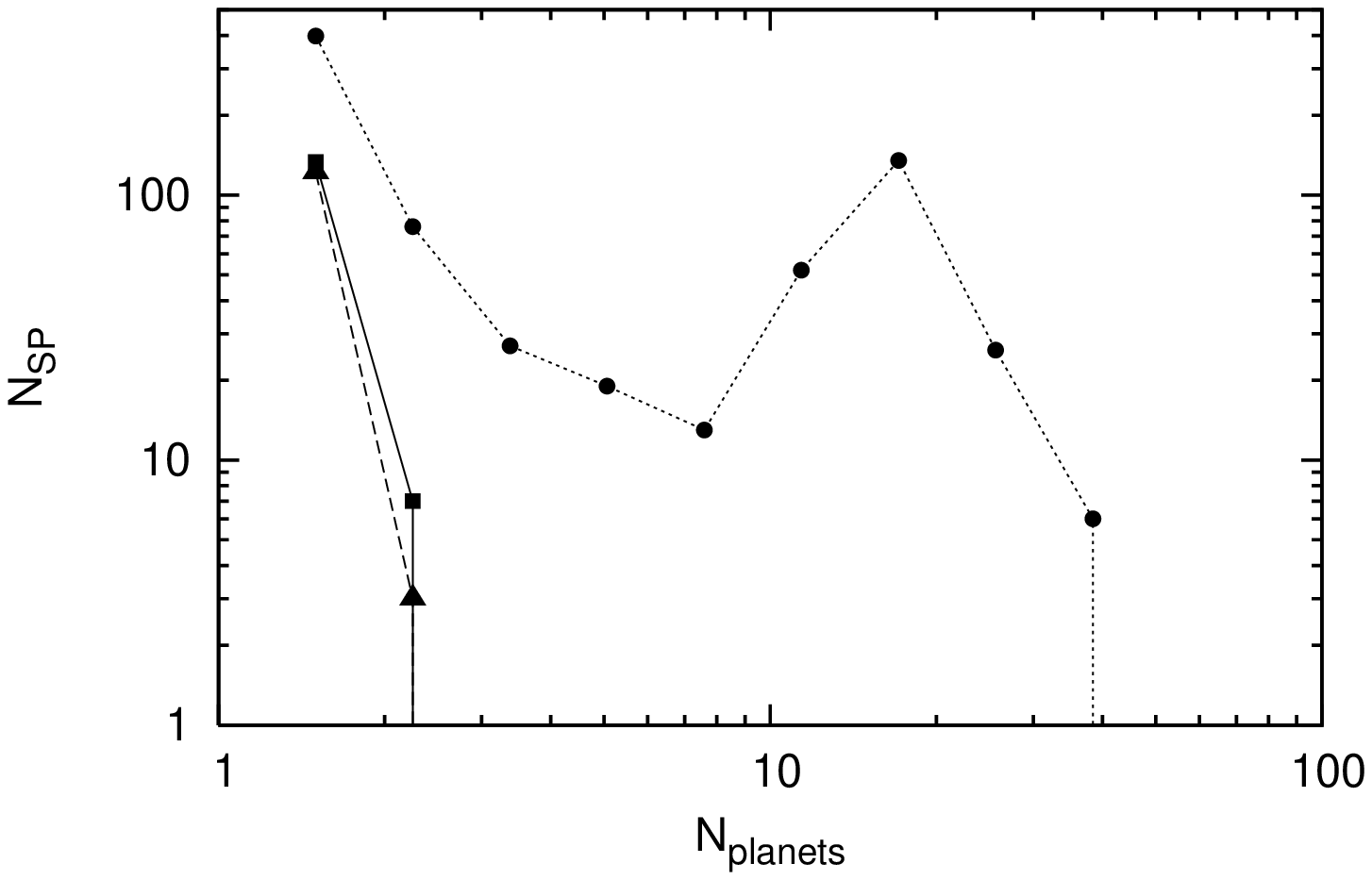}}
    \subfigure[]{\label{fig:histon-c01}\includegraphics[angle=270,width=.44\textwidth]{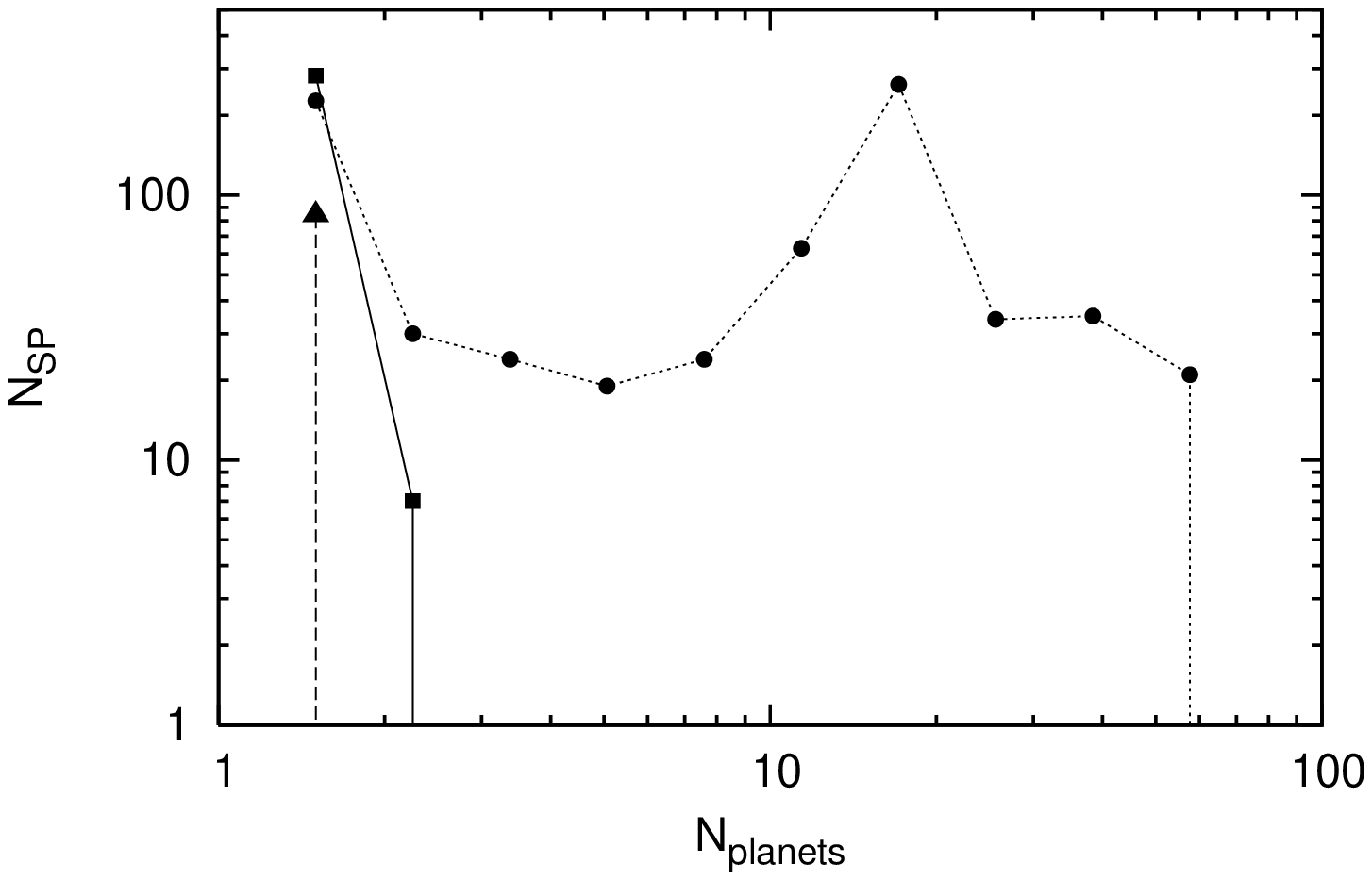}} 
    \subfigure[]{\label{fig:histon-c001}\includegraphics[angle=270,width=.44\textwidth]{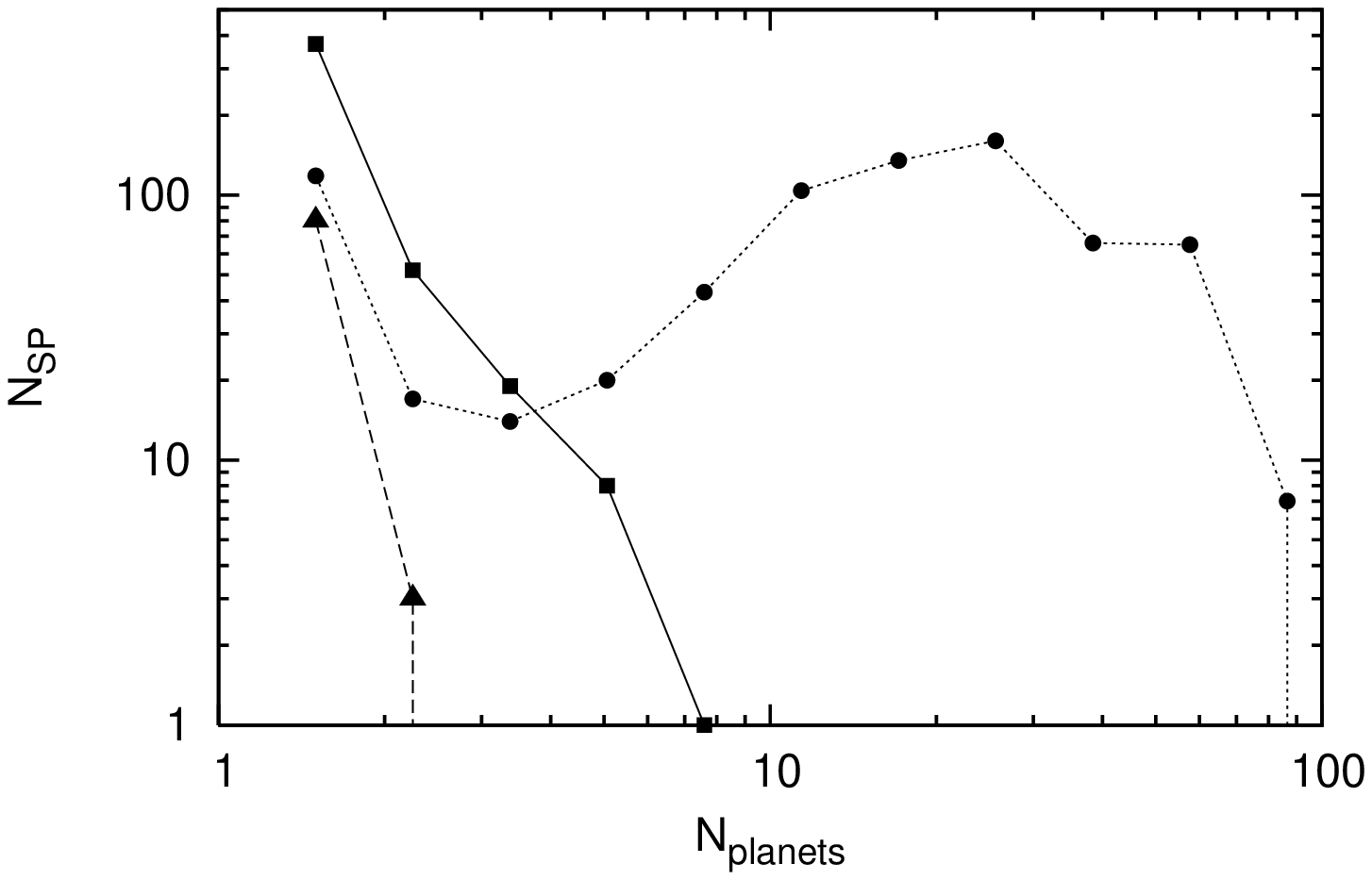}}
  \end{center}
\caption{Histogram of the final number of terrestrial (dotted line with filled circles), giant with low percentage of  gas (dashed line with filled triangles) and gas giant planets (solid line with squares) per planetary disc. This histograms were obtained with different values of $C_{migI}$. In Fig. \ref{fig:histon-c1} $C_{migI}=1$, figure \ref{fig:histon-c01} was obtained when $C_{migI}=0.1$ and in figure \ref{fig:histon-c001} the value of $C_{migI}$ is 0.01. The number of planetary systems with few terrestrial planets are larger when $C_{mig}$ is 1, and it decreases when the type I migration rate slows down. On the contrary, the number of planetary systems with few gas giants are smaller with $C_{migI}=1$, and it increases with small values of $C_{migI}$.}
\label{histon}
\end{figure}

We also analyse the kind of planets formed per disc. We consider that terrestrial planets are those with $M_t<7M_{\oplus}$,  giants with low percentage of  gas  are suppose to be the planets with $M_t>7M_{\oplus}$ and a percentage of gas mass less than $<15 \%$, and the others, which have a larger percentage of gas are gas giants. The histogram presented in Fig. \ref{histon} shows the number of terrestrial planets (dotted line with filled circles), giants with low percentage of  gas (dashed line with filled triangles) and gas giants (solid line with squares) formed per planetary system.

Figure \ref{fig:histon-c1} is the one obtained when $C_{migI}=1$. We can see that planetary systems with few planets have mostly terrestrial ones and there are few with giant planets. As we explained, for this rapid migration rate, the range of discs which form planetary systems with low $N_{planets}$ are those with $0.5<f_d<30$. In this range most of the discs have low mass and cores have less time to grow because they are rapidly moved to the central star. As a result most of the planetary systems with few planets will form only terrestrial planets. On the other hand those discs with a large value of $f_d$ are able to form planetary systems with giant planets, but are the less ones. We also note a big absence of planetary systems with an intermediate number of planets ($N_{planets}>30$), this is because of the rapid migration rate, which moves quickly the cores to the inner disc limit, so they accrete the other cores to its path, forming planetary systems with few planets.

When $C_{migI}$ is smaller (Fig. \ref{fig:histon-c01} with $C_{migI}=0.1$ and \ref{fig:histon-c001} with $C_{migI}=0.01$), the population of planetary systems with few gas giant planets increases, meaning that the formation of gas giants is favoured with slower type I migration rates. On the other hand, the number of giant planets with low percentage of gas remains low, this population is not affected by the rate of migration. We also note that the number of planetary systems with an intermediate number of planets increases, so when migration slows down the final amount of planets per disc tends to follow a uniform distribution.

\subsection{Mass and semimajor axis distribution}

To investigate the mass and semimajor axis of extrasolar planets we perform a series of numerical simulations considering different prescriptions for the gas accretion rate and different retardation constants for type I migration. 

\begin{figure}
  \begin{center}
    \subfigure[]{\label{fig:NMc1}\includegraphics[angle=270,width=.44\textwidth]{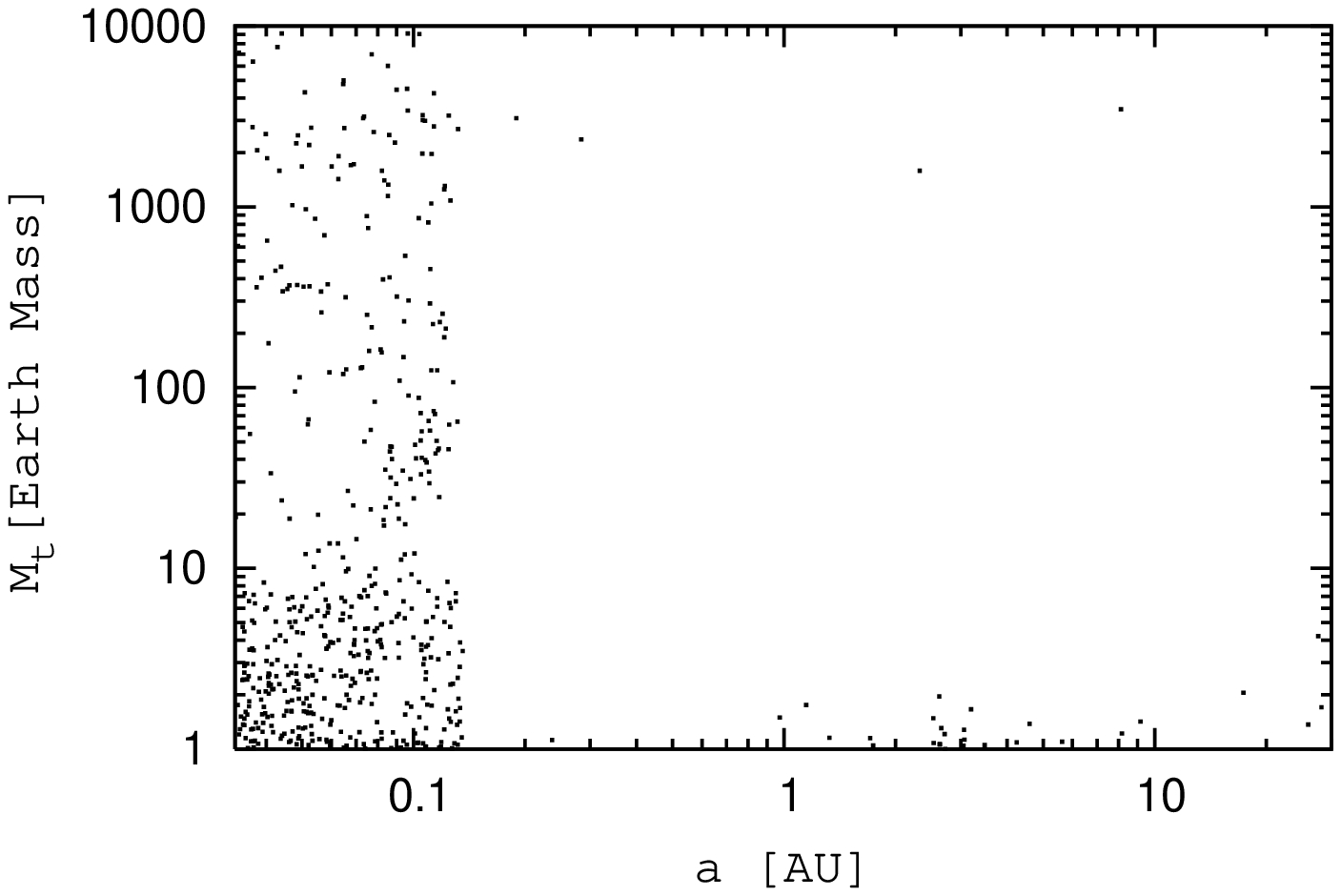}}
    \subfigure[]{\label{fig:NMc01}\includegraphics[angle=270,width=.44\textwidth]{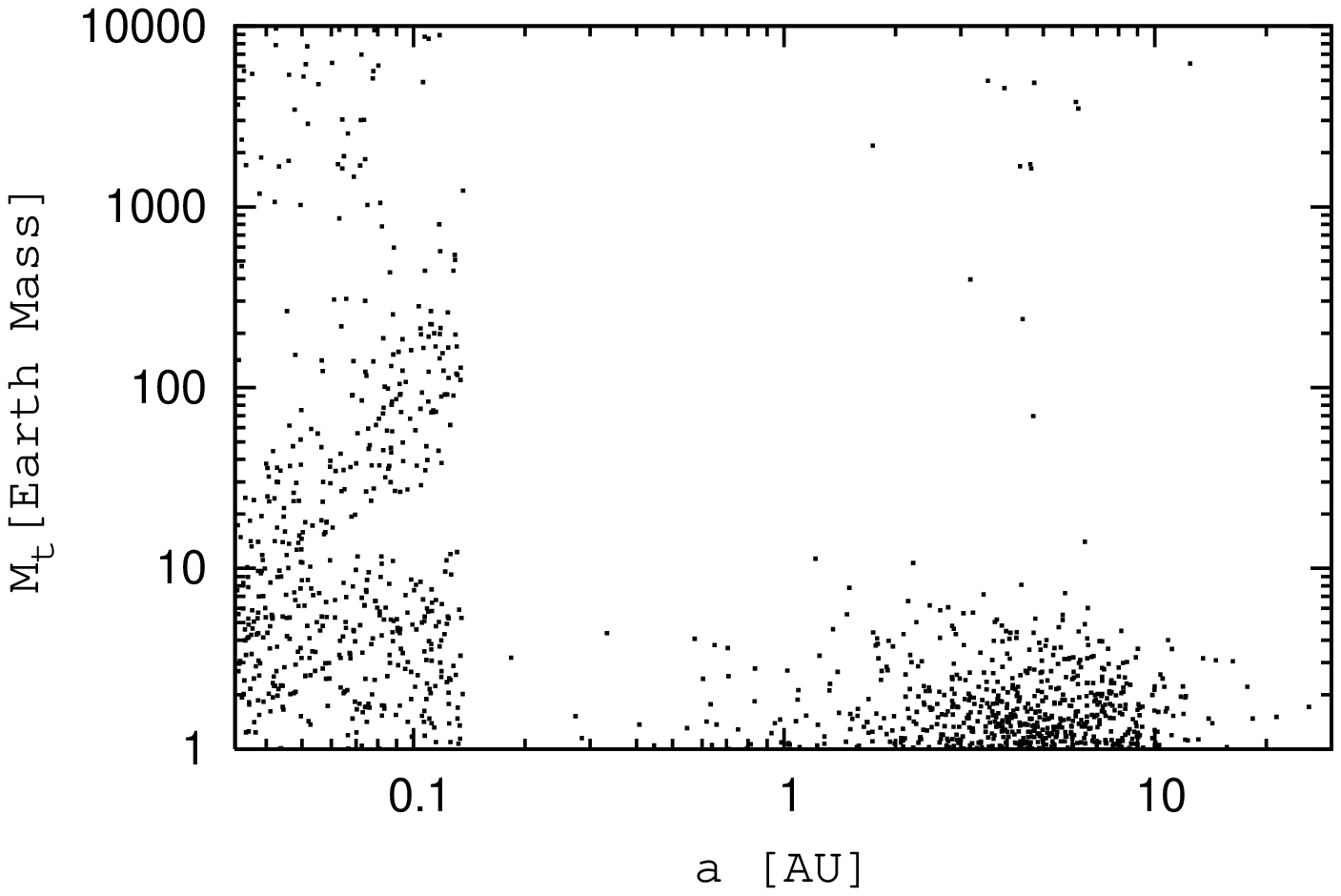}}
    \subfigure[]{\label{fig:NMc001}\includegraphics[angle=270,width=.44\textwidth]{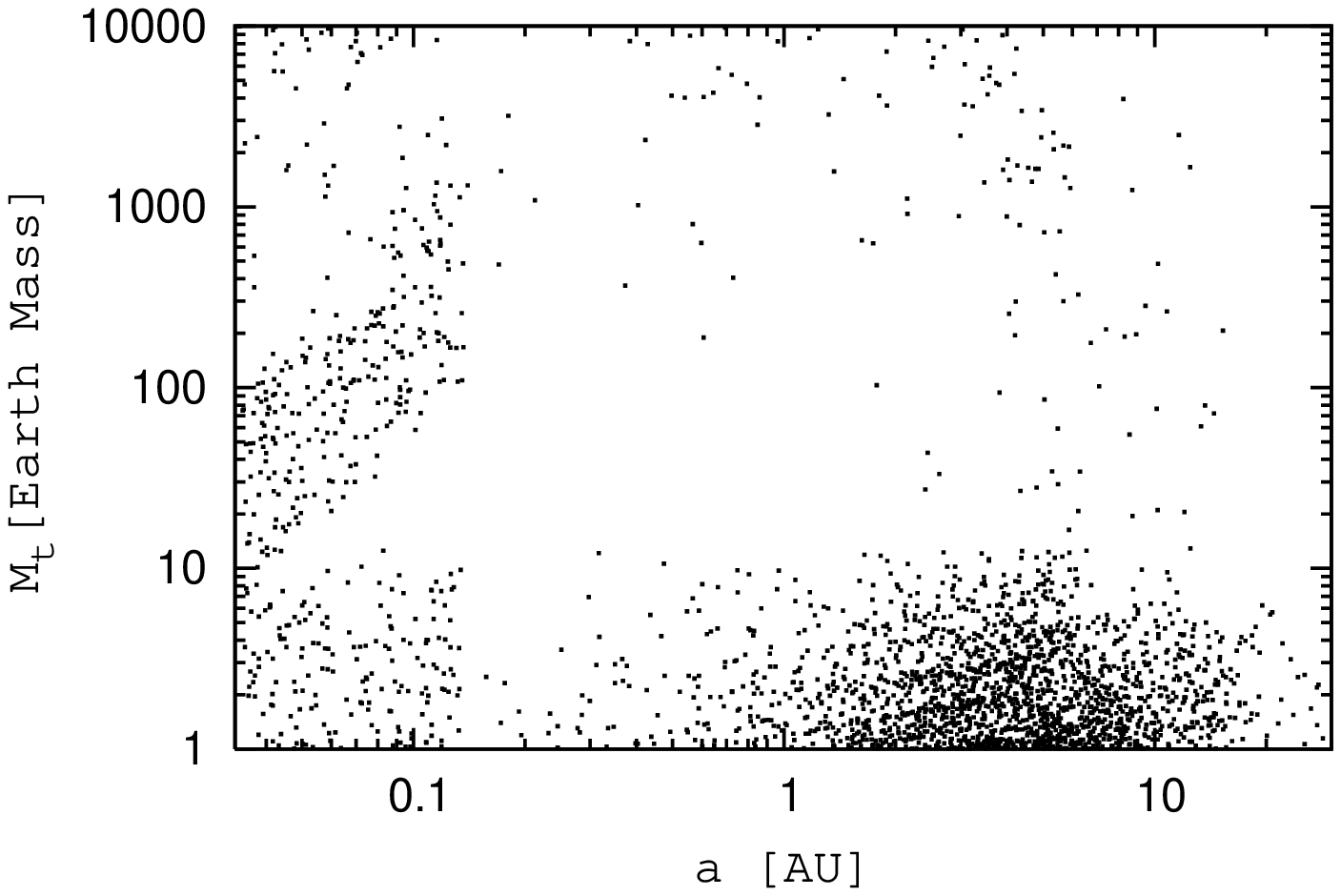}} 
    \subfigure[]{\label{fig:NMc0}\includegraphics[angle=270,width=.44\textwidth]{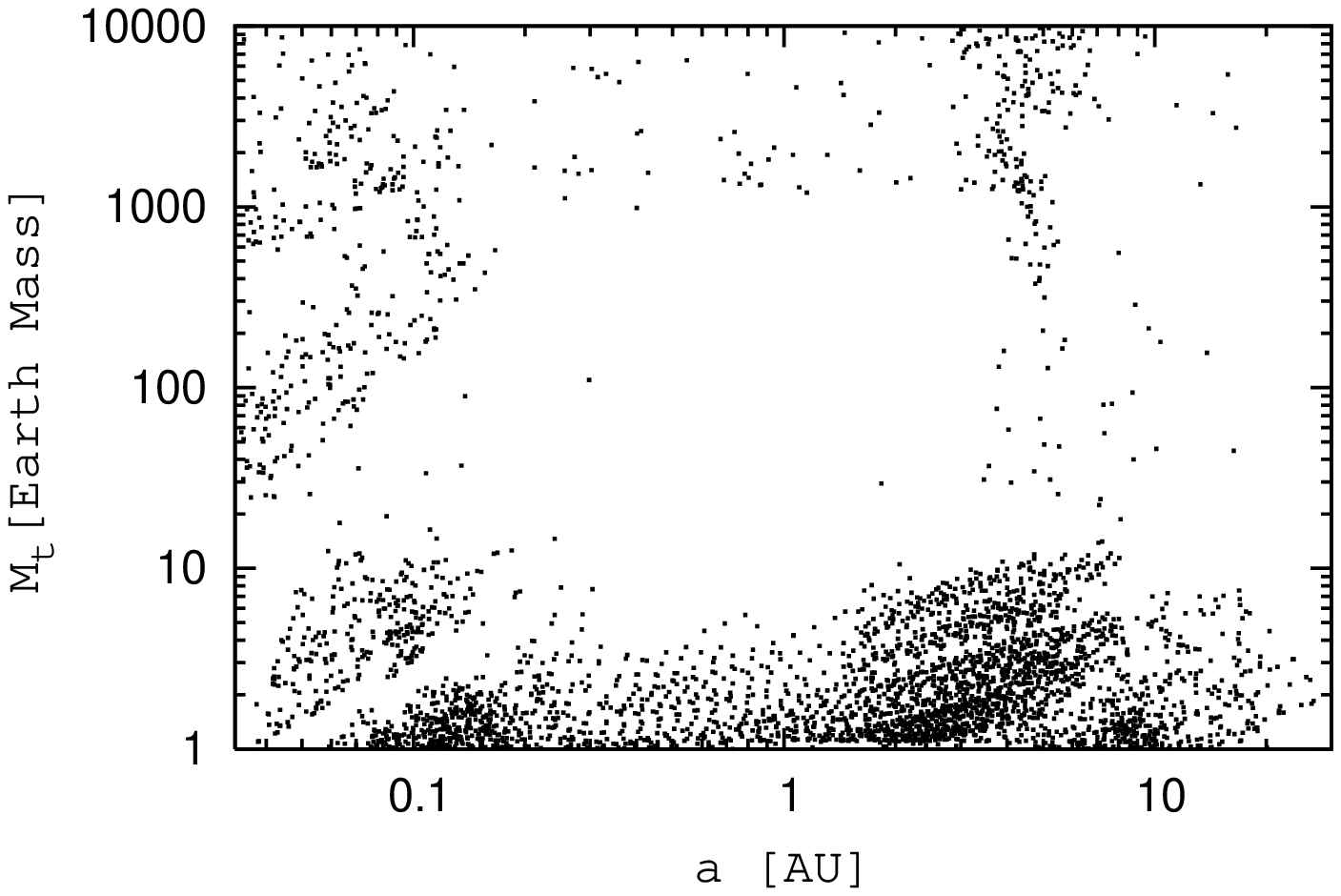}} 
  \end{center}
  \caption{Distribution of masses and semimajor axis obtained with our model. Figure \ref{fig:NMc1} shows the data obtained with $C_1=1$, in  \ref{fig:NMc01} $C_1=0.1$, in  \ref{fig:NMc001}  $C_1=0.01$ and the effect of type I migration is not considered in  \ref{fig:NMc0}. We find a larger population of terrestrial planets and a lower population of gas giant beyond the snow line than those observed on Fig. \ref{IL}. We also observed a distribution more populated as migration slows down ($a\to d$).}
  \label{NM}
\end{figure}

\begin{figure}
  \begin{center}
    \subfigure[]{\label{fig:ILc1}\includegraphics[angle=270,width=.44\textwidth]{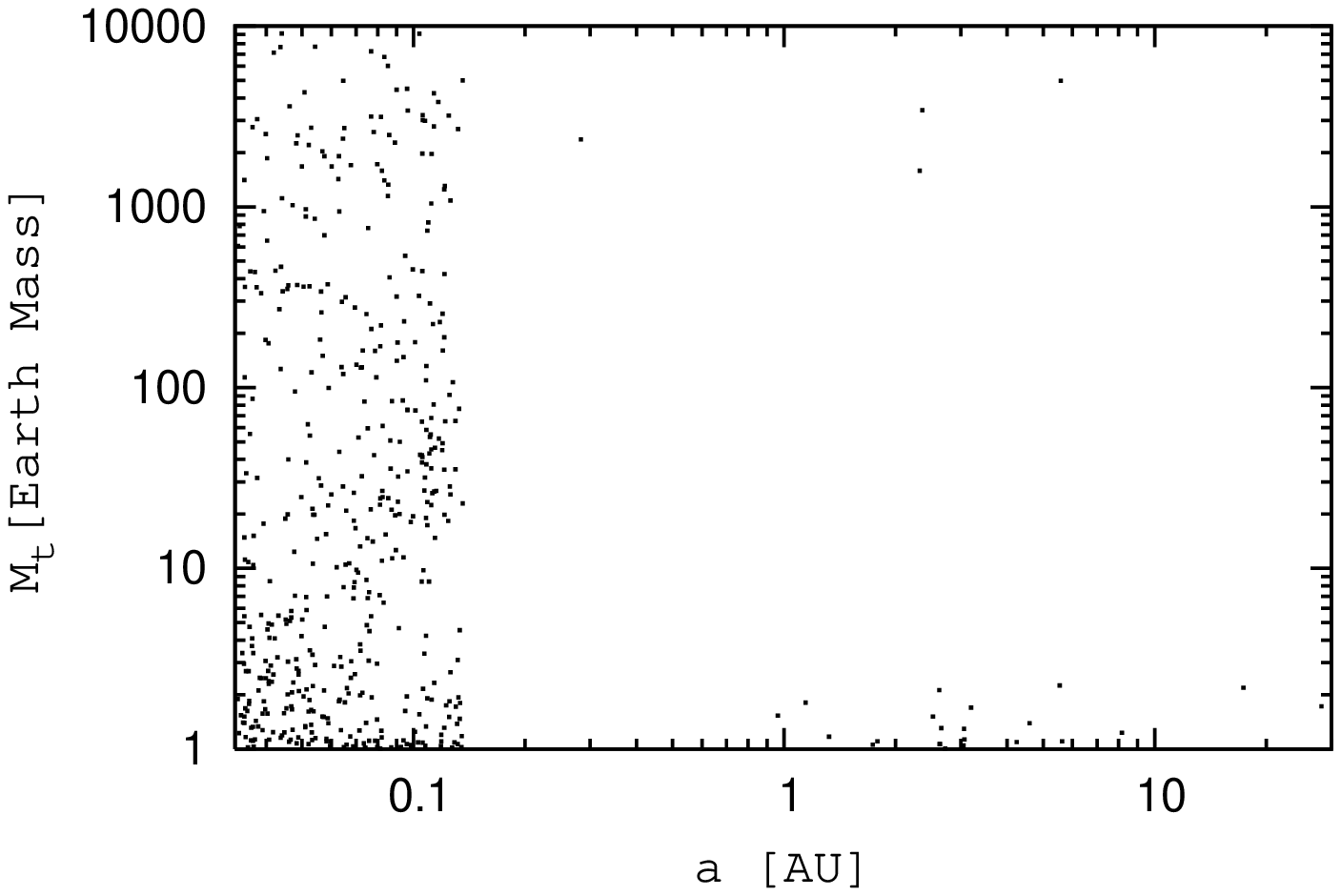}}
    \subfigure[]{\label{fig:ILc01}\includegraphics[angle=270,width=.44\textwidth]{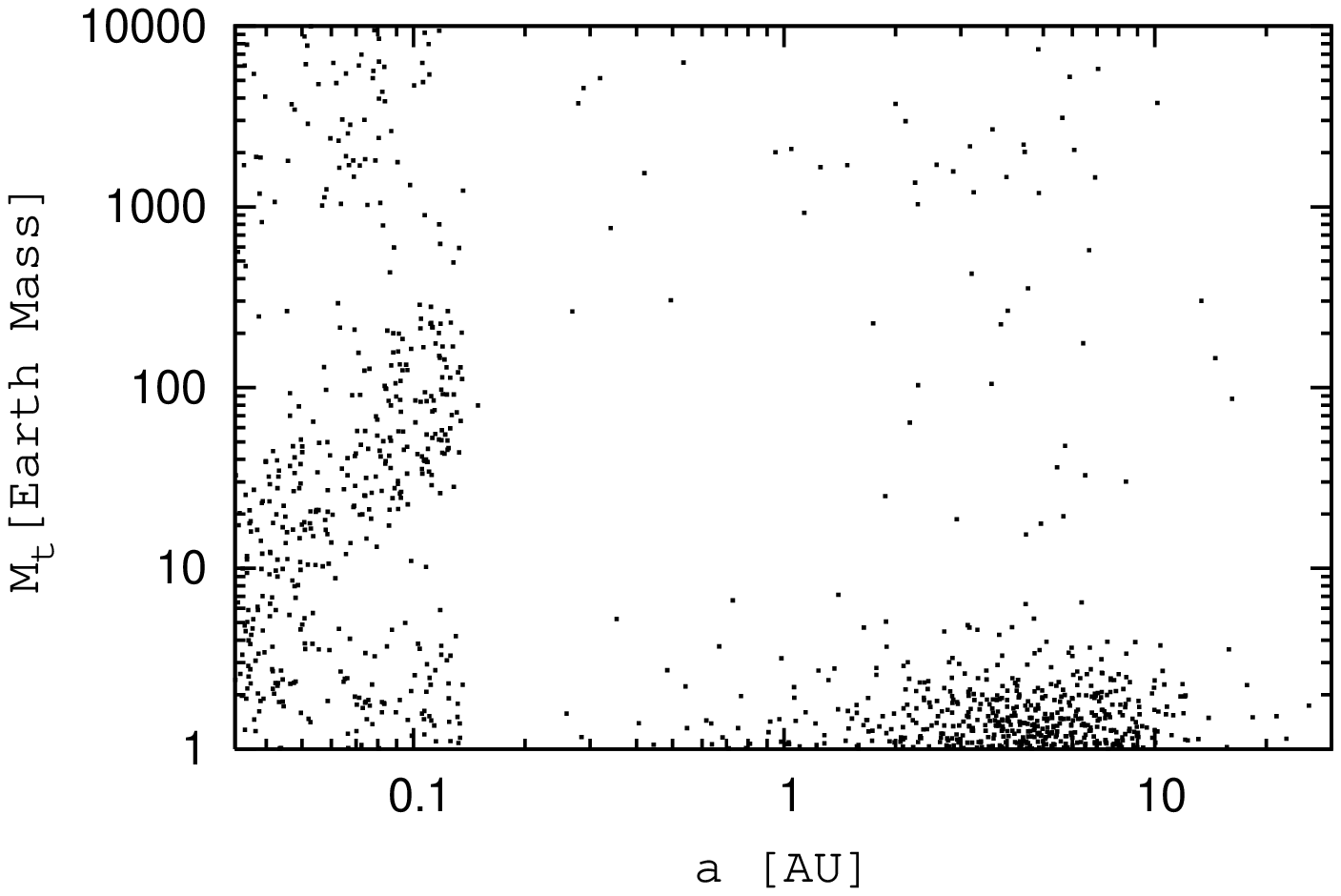}} 
    \subfigure[]{\label{fig:ILc001}\includegraphics[angle=270,width=.44\textwidth]{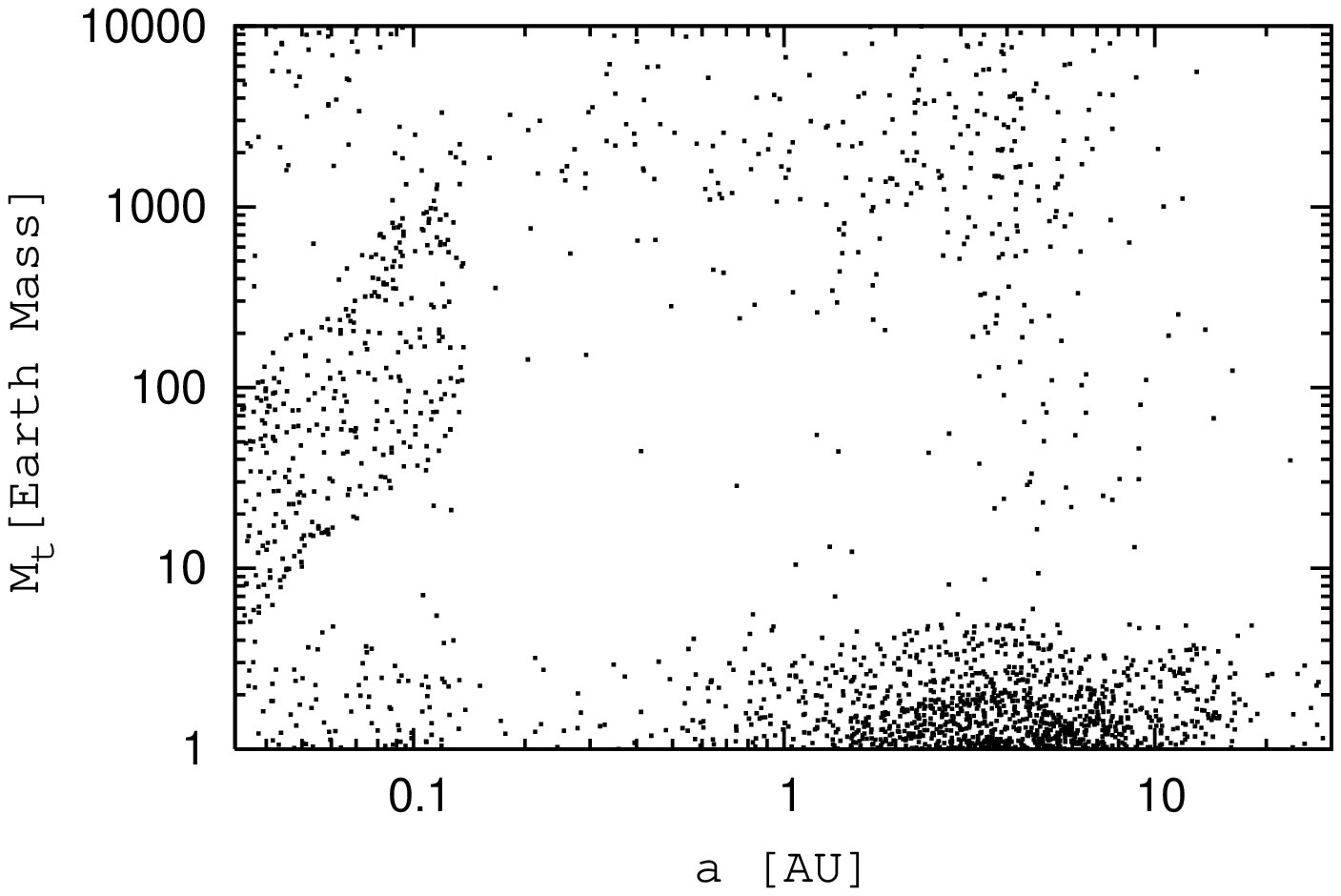}}
  \subfigure[]{\label{fig:ILc0}\includegraphics[angle=270,width=.44\textwidth]{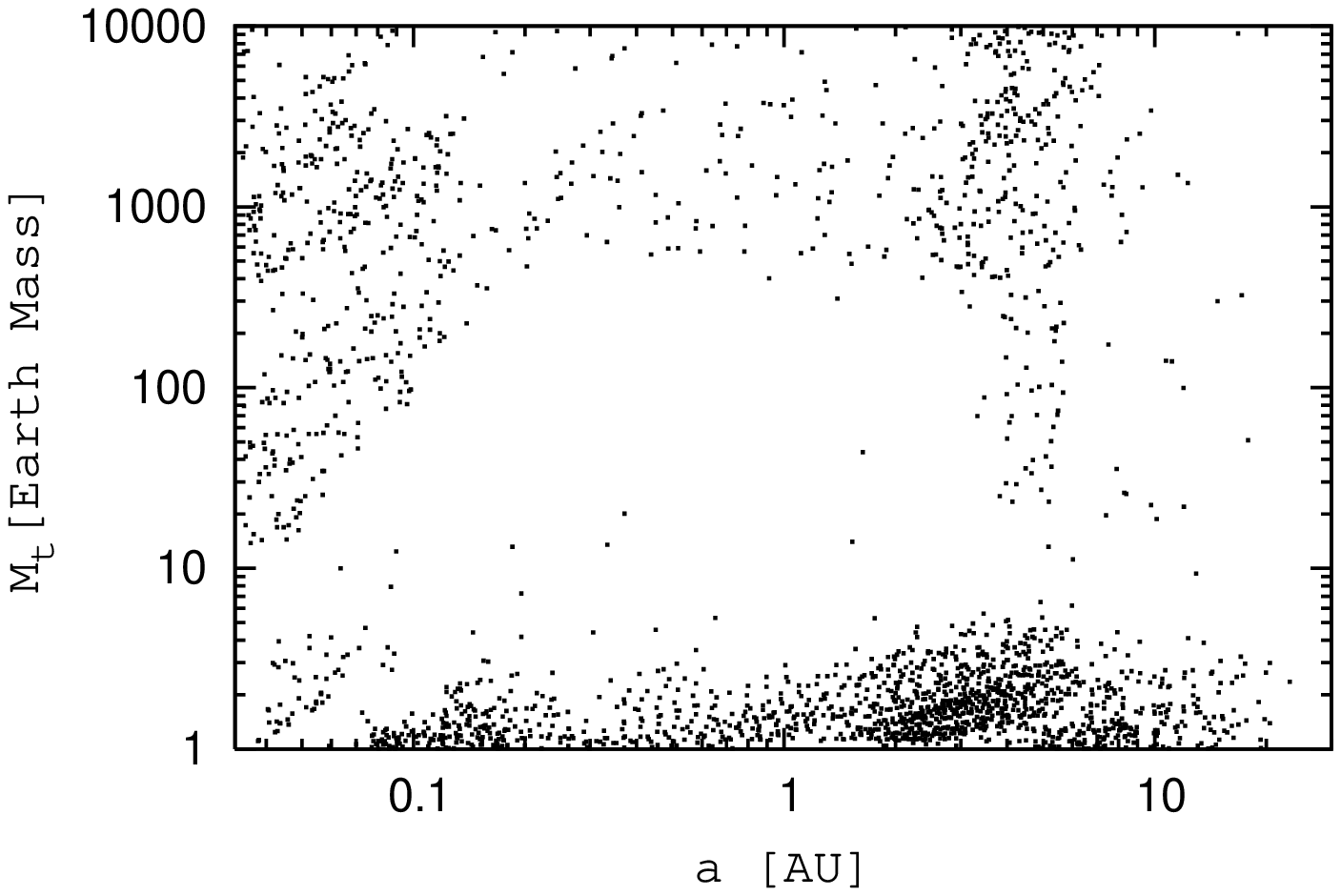}}
  \end{center}
\caption{Similar figures as \ref{NM}, but the gas accretion rates of IL04 are considered. A comparison with Fig. \ref{NM}, shows a smaller population of terrestrial planets and a larger  of giant planets with $M_t \ge 100 M_{\oplus}$, especially when the migration slows down,\ref{fig:ILc01}, \ref{fig:ILc001} and \ref{fig:ILc0}.}
\label{IL}
\end{figure}

Results are displayed in Fig. \ref{NM}(a,b,c,d) and \ref{IL}(a,b,c,d). The first ones were found by considering the gas accretion rates obtained by fitting the results of \citet{b6} and the second ones were obtained using the gas accretion rates given by IL04. In the figures \ref{fig:NMc1} and \ref{fig:ILc1} the factor considered for delaying the type I migration rate is $C_{migI}=1$, in Fig.\ref{fig:NMc01} and \ref{fig:ILc01} $C_{migI}=0.1$ Fig. \ref{fig:NMc001} and \ref{fig:ILc001} are those obtained with $C_{migI}=0.01$ and finally Fig. \ref{fig:NMc0} and \ref{fig:ILc0} where found without considering the effects of type I migration ($C_{migI}=0$).

As Fig. \ref{NM} and \ref{IL} show, the planetary distribution of mass and semimajor axis is strongly dependent on the gas accretion model considered (Paper I), as well as the type I migration rate used.

When $C_{migI}=1$ most of the planets migrate to the inner disc radius. The mainly difference between the distribution found with our model and the one obtained with IL04's gas accretion rate (Fig. \ref{fig:NMc1}and \ref{fig:ILc1},respectively), is the population of planets with masses between $1$ and $10 M_{\oplus}$, which is larger on the first one.

 In Fig.\ref{fig:NMc01} and \ref{fig:ILc01} ($C_{migI}=0.1$), the type I migration is slower and there are less planets who reach the inner edge of the disc. We can see a major population of planets with $1-10 M_{\oplus}$ and $a\ge 1au$. The population of giant planets are also increased, but this is larger in Fig.\ref{fig:ILc01} because with IL04's gas accretion rate the run away of gas is reached sooner, and also the $M_{gap}$. This allows the planets to start the type II migration, which is slower than the type I.

Type I migration is the slowest when $C_{migI}=0.01$, and as seen in Fig.\ref{fig:NMc001},\ref{fig:ILc001},\ref{fig:NMc0} and \ref{fig:ILc0}, the distribution found when $C_{migI}=0.01$ is very similar to those found with type II migration only. When $C_{migI}=0.01$ the population of planets with masses less than 10 $M_{\oplus}$ is larger than in the other cases and there are more giant planets, specially when $a\ge 1au$. The differences between \ref{fig:NMc001} and  \ref{fig:ILc001} are similar to those found when $C_{migI}=0.1$, which is in agreement with the results found in Paper I.

The ``planetary desert'', which is a region with a deficit of planets, was found to be between $100-1000 M_{\oplus}$ (Paper I) with our model. Here we found that the effect of the planetary migration and specially type I migration, is to enlarge the desert, allowing to the planets with masses between $10$ and $100M_{\oplus}$ to reach the inner limit of the disc and empty the area with semimajor axis between $\approx 0.2$ and $\approx 3 au$.



Another important result is that we found a major population of habitable planets with our model than the one found with IL04's gas accretion rates. In order to compare the different population of terrestrial planets found with different gas accretion rates, we adopt a criteria where a habitable planet is a planet with masses between $\approx 0.3$ and $\approx 7M_{\oplus}$ \citep{b15}. The upper limit to the mass was obtained by \citet{b15} with a different time-scale for gas accretion, with our time-scale this value is probably larger, but we consider $7 M_{\oplus}$ as a nominal value for comparison.

 The Table \ref{terrestres} shows the percentage of final planets that are habitable according to \citet{b15}'s definition. In the table, model I is the one considering our gas accretion rate, and model II consider IL04's gas accretion rate. 

\begin{table}
\centering
 \caption{Percentage of final planets that are habitable ($0.3 \le M_t \le 7M_{\oplus}$ according to \citet{b15}'s criteria) found with our model (Model I), compared with the results obtained with IL04's gas accretion rate (Model II). Different delaying type I migration constants are considered.}
\begin{tabular}{|p{4cm}|p{2.7cm}|}
 \hline
 Scenario & Habitable Planets ($0.3 \le M_t \le 7M_{\oplus}$) \\
\hline
\hline
Model I, $C_{migI}=1$ & 87.02 (\%) \\

Model II, $C_{migI}=1$ & 82.2 (\%) \\
\hline
Model I, $C_{migI}=0.1$ & 90.5 (\%) \\

Model II, $C_{migI}=0.1$ & 86.9 (\%) \\
\hline
Model I, $C_{migI}=0.01$ & 91.57 (\%) \\

Model II, $C_{migI}=0.01$ & 87.18 (\%) \\
\hline
\label{terrestres}
\end{tabular}
\end{table}

With our model the run away gas accretion process is reached by the cores at a larger mass (Paper I) so there are terrestrial planets that never reach this ``crossover mass'' and for this reason we found more planets with masses less than $7 M_{\oplus}$. So a slower gas accretion rate leads to larger population of habitable planets.

\section{Conclusions}

In our previous work, we had developed a very simple model for computing planetary formation, which had allowed us to show the strong dependence between the gas accretion model considered and the mass distribution of extrasolar planets.

In order to get a better understanding of the processes involved in planetary formation, we improve our previous model, including significant effects that had not been taken into account in Paper I while maintaining the computational 
speed.

We have presented the results of the simulations performed with our improved model, considering the formation of several planets per disc and taking into account the collision among them as a source of potential growth. The formation of several cores simultaneously in the disc has a strong influence on the dynamics of the planetesimal disc. The evolution due to this effect and due to the gas drag effect was also considered in this work. 

Finally, we have analysed the interaction between the protoplanets and the disc, which leads to a planetary migration (type I and II). The migration of a giant planet towards the central star could perturbate the cores placed in its path, causing the ejection or the accretion of  the core by the giant planet. The another possibility is the survival of the core to the passage of the giant planet. We have performed numerical simulations taking into account that in a close encounter a terrestrial planet has the 71 percent chance of surviving the passage of a giant, but these simulations have not produced significant changes in the results due to the fact that in a real system several protogiant planets cross the region of terrestrial planets.

The final number of planets formed per disc is a direct result of the planetary system evolution. We showed that the distribution is strongly dependent on the type I migration rate considered, noting an absence of planetary systems with an intermediate number of planets when the fastest migration rate is considered and seeing that it becomes more uniform when the migration slowed down. 

The final number of planets per planetary system also depends on the initial mass of the disc, those discs with low mass will form planetary systems with a large number of planets and a higher mass disc is necessary to form planetary systems with a few ones, but the range of disc masses corresponding to the final number of planets depends on the type I migration rate considered. For a rapid migration rate low mass discs are able to form planetary systems with few planets, but when the migration slows down only high mass discs could form this kind of planetary systems. 

However we note that many of the systems with a large number of planets will be unstable and undergo further evolution since the orbital spacing is small, and eccentricities will increase after the gas is gone. The final number of planets will be substantially smaller in some systems, but we do not consider this fact in this work.

We also have analysed the kind of planets formed per disc, and found that planetary systems with a large number of planets have only terrestrial ones, but when we looked at those planetary systems with few planets, the kind of planets formed varies according to the type I migration rate considered. When we used the fastest rate, most of the planets are terrestrial planets, but when the rate is slower the amount of terrestrial planets decreases and most are now giants. In conclusion we found that the gas giant formation is favored at slower type I migration rate. On the other hand the giant planets with few gas reminds little in all the simulations, this population is not affected by the rate of migration.

When we analyse the mass and semimajor axis distribution we consider different gas accretion rates, and the results show that this distribution is strongly dependent on the gas accretion model and also on the rate of migration, mainly due to type I migration effects. The boundaries of the planetary desert are enlarged due to the rapid migration of the embryos that did not reach the necessary mass to open up a gap. But we note that the planetary desert would be a bit different, since the actual observational sample shows that the number of observed planets at distances smaller than 0.1 AU from the central star is not as large as that found through numerical simulations, which could mean that most planets do not stop their migration at the inner edge of the disc. Finally we found that a lower gas accretion rate leads to a larger population of habitable planets, and lower population of gas giants beyond the ice condensation line.

\end{document}